\documentclass{moriond}

\def\be{\begin{equation}}
\def\ee{\end{equation}}
\def\bea{\begin{eqnarray}}
\def\eea{\end{eqnarray}}

\newcommand{\Photo}{\includegraphics[height=35mm]{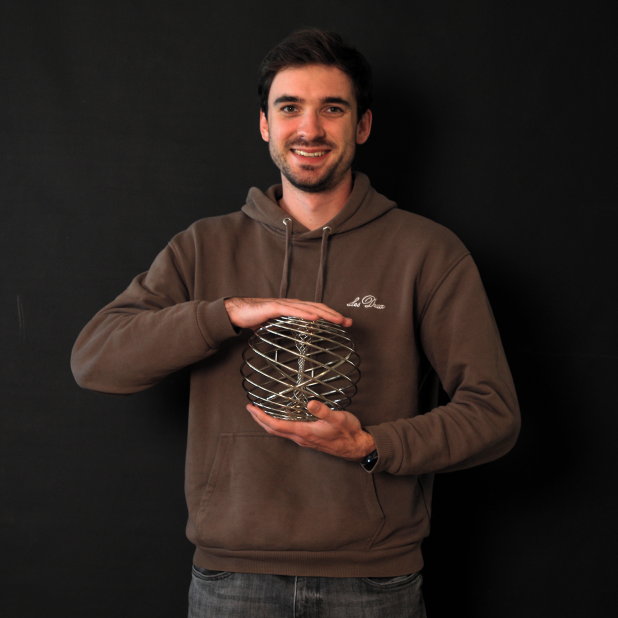}}

\begin{document}
\vspace*{4cm}
\title{Standard Model W, Z (+jet) at CMS and ATLAS}

\author{ C. Verstege on behalf of the ATLAS and CMS Collaborations }

\address{Karlsruhe Institute of Technology, Institute of Experimental Particle Physics, Wolfgang-Gaede-Str. 1,\\
76131 Karlsruhe, Germany}

\maketitle\abstracts{
Recent measurements of Standard Model W and Z boson production performed by the ATLAS and CMS Collaborations at the CERN Large Hadron Collider are reviewed.
These include a search for charged lepton flavour violating decays of Z and Z' bosons, a measurement of W-boson angular coefficients and transverse momentum distributions, a triple-differential measurement of Z+jet production, and a study of the jet mass in boosted hadronic W boson decays with an extraction of the W boson mass.
Together, these results highlight the role of precision measurements as stringent tests of perturbative quantum chromodynamics and electroweak theory.
The large Run~2 dataset enables multi-differential measurements with enhanced sensitivity to parton distribution functions and increasingly precise tests of theoretical predictions.
}

\section{Introduction}
In proton-proton collisions at the CERN Large Hadron Collider (LHC), W and Z bosons are produced with large cross sections, providing experimentally clean and theoretically well-understood processes.
Precision measurements of their production and decay properties constitute essential tests of perturbative quantum chromodynamics (QCD) and electroweak (EW) theory.

In addition, these processes are sensitive to parton distribution functions (PDFs), providing valuable input for improving the theoretical description of proton--proton collisions.
Measurements performed in association with jets further probe higher-order QCD effects.
The large datasets collected during Run~2 enable multi-differential measurements with unprecedented precision, allowing detailed tests of theoretical predictions across a wide kinematic range.

This contribution \footnote{Copyright 2026 CERN for the benefit of the ATLAS and CMS Collaborations. Reproduction of this article or parts of it is allowed as specified in the CC-BY-4.0 license.} summarizes recent results from the ATLAS~\cite{ATLAS:2008xda} and CMS~\cite{CMS:2008xjf,CMS:PRF-21-001} Collaborations, focusing on W and Z boson production, including associated jet activity and boosted topologies.
Leptonic decay channels are primarily used due to their clean experimental signatures and strong suppression of QCD multijet backgrounds, enabling high-precision measurements.

\section{Search for charged lepton flavour violating Z and Z' boson decays}

A search for charged lepton flavour violating (CLFV) decays of the Z boson was performed by the CMS Collaboration~\cite{CMS:SMP-23-003} using proton-proton collision data at $\sqrt{s}=13\,\mathrm{TeV}$, corresponding to an integrated luminosity of $138\,\mathrm{fb}^{-1}$.
In the Standard Model (SM), CLFV processes are highly suppressed, with branching fractions of $\mathcal{O}(10^{-60})$ when extended to include neutrino masses, making any observable signal a clear indication of physics beyond the Standard Model.

The analysis targeted the decay channels $Z \to e\mu$, $Z \to e\tau$, and $Z \to \mu\tau$, as well as potential heavy neutral resonances ($Z'$) decaying to $e\mu$ in the mass range $110$--$500\,\mathrm{GeV}$.
Signal extraction was based on the invariant mass of the dilepton system and multivariate discriminants.
In the $e\mu$ channel, where the Z boson is fully reconstructed, the signal yield was obtained from fits to the invariant mass spectrum.
For the $e\tau$ and $\mu\tau$ channels, boosted decision tree (BDT) discriminants were used as the primary observables due to the reduced mass resolution from the presence of neutrinos in the $\tau$ decay.

No significant deviation from the SM expectation was observed.
Upper limits at $95\%$ confidence level were set on the branching fractions:
\[
\begin{array}{rcl}
\mathcal{B}(Z \to e\mu) & < & 1.9 \times 10^{-7}, \\
\mathcal{B}(Z \to e\tau) & < & 1.38 \times 10^{-5}, \\
\mathcal{B}(Z \to \mu\tau) & < & 1.20 \times 10^{-5}.
\end{array}
\]
The limit on $\mathcal{B}(Z \to e\mu)$ is the most stringent direct constraint to date.
In addition, limits on $\sigma(\mathrm{pp} \to Z') \times \mathcal{B}(Z' \to e\mu)$ were set in the range $0.3$--$7\,\mathrm{fb}$ for $Z'$ masses between $110$ and $500\,\mathrm{GeV}$.

These results significantly constrain a broad class of beyond-the-Standard-Model scenarios predicting CLFV and demonstrate that the sensitivity is currently limited primarily by the available data sample.

\section{Measurement of W-boson angular coefficients and transverse momentum}

The first measurement of the full set of angular coefficients describing W-boson production in proton--proton collisions at $\sqrt{s}=13\,\mathrm{TeV}$ was presented by the ATLAS Collaboration~\cite{ATLAS:2025mlt}.
The analysis used a dedicated low pile-up dataset corresponding to an integrated luminosity of $338\,\mathrm{pb}^{-1}$, which enabled an improved reconstruction of the hadronic recoil and, consequently, of the W-boson transverse momentum.

The differential cross section for W-boson production and decay can be expressed as a decomposition into harmonic polynomials of the lepton decay angles, multiplied by angular coefficients $A_i$ that encode the helicity structure of the process.
These coefficients are sensitive to QCD effects in vector-boson production and provide direct information on the W-boson polarisation.

The measurement was performed in the full phase space of the decay leptons for both $W^+$ and $W^-$ bosons, and extracted the eight angular coefficients together with the differential cross section as a function of the W-boson transverse momentum $p_T^W$.
A profile likelihood fit was used to determine the coefficients in bins of $p_T^W$.

Several angular coefficients, in particular $A_0$ and $A_{2-4}$, were found to be significantly different from zero, demonstrating the sensitivity of the measurement to the underlying QCD dynamics.
The results were compared to state-of-the-art predictions at next-to-next-to-leading order (NNLO) in QCD, including resummation effects, and showed overall good agreement.

\section{Triple-differential measurement of Z+jet production}

The CMS Collaboration reported the first triple-differential measurement of Z boson production in association with at least one jet~\cite{CMS:SMP-24-010} in proton--proton collisions at $\sqrt{s}=13\,\mathrm{TeV}$, using a dataset corresponding to an integrated luminosity of $138\,\mathrm{fb}^{-1}$.

\begin{figure}
    \centering
    \includegraphics[width=0.45\linewidth]{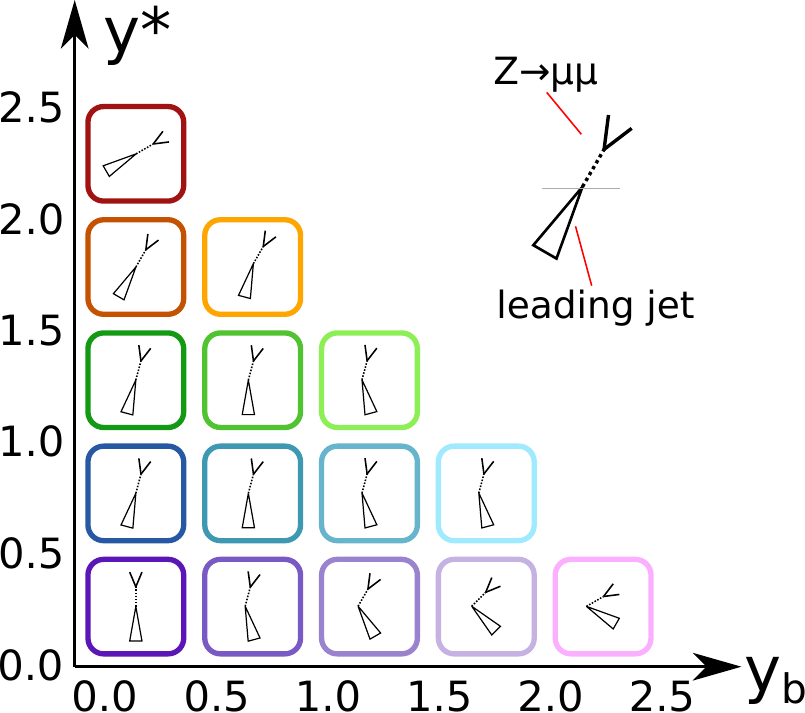}
    \includegraphics[width=0.45\linewidth, trim={0 1cm 0 1cm}]{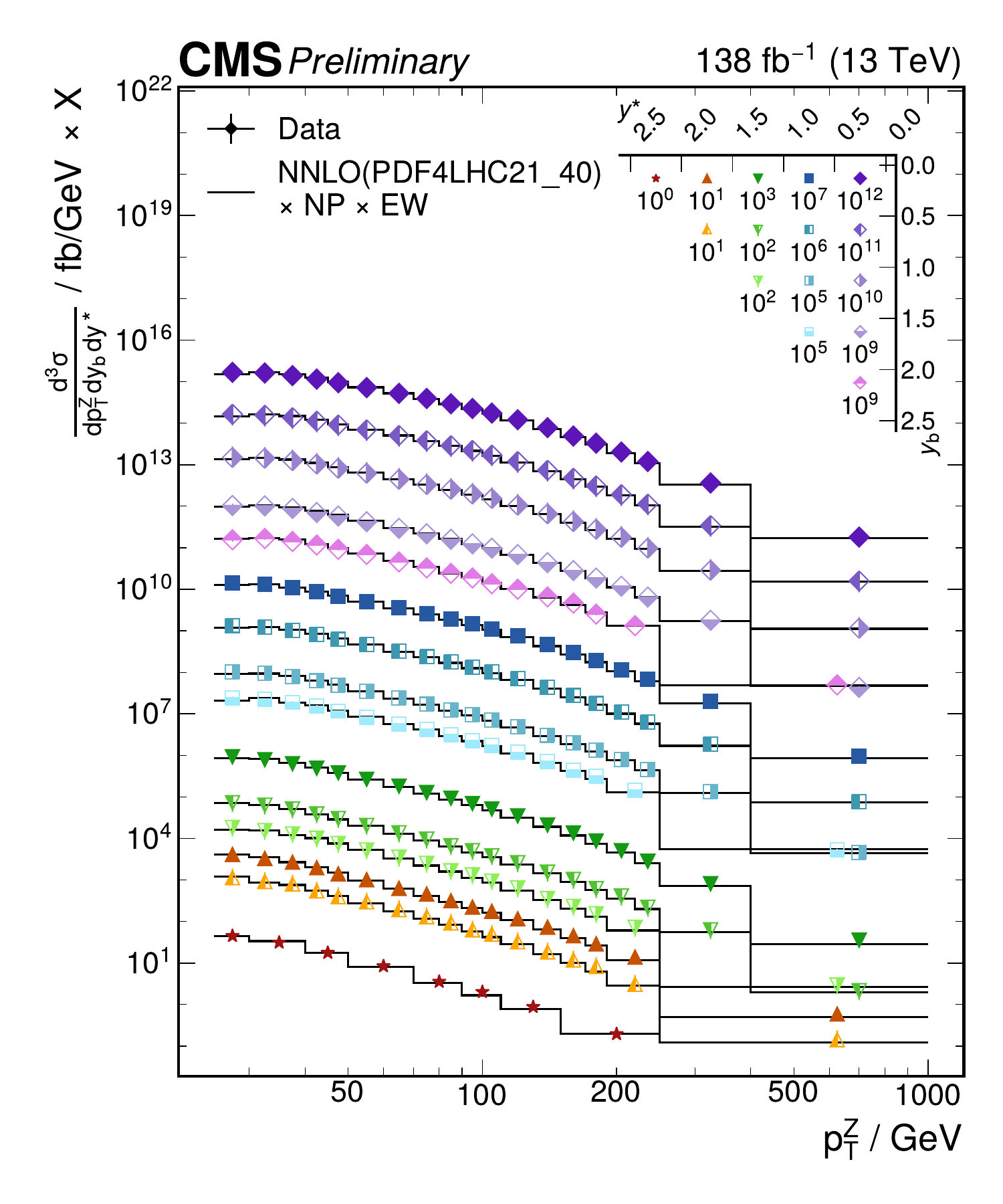}
    \caption{Illustration of the binning in $y_b$ and $y^*$ for the triple-differential Z+jet cross section measurement (left). Data compared to theory predictions at NNLO in QCD, corrected for non-perturbative and electroweak effects, are shown on the right for all bins in $y_b$ and $y^*$ as a function of $p_T^Z$.~\protect\cite{CMS:SMP-24-010}}
    \label{fig:3dzjet}
\end{figure}

The cross section was measured as a function of the Z boson transverse momentum $p_T^Z$, half of the rapidity separation between the Z boson and the leading jet $y^*$, and the boost of the Z+jet system $y_b$.
These observables are directly related to the kinematics of the underlying $2 \to 2$ scattering process: $p_T^Z$ sets the energy scale, while $y^*$ and $y_b$ probe the scattering angle and the momentum fractions of the incoming partons, respectively.
An illustration of the final-state kinematics and the binning in $y^*$ and $y_b$ is shown in Fig.~\ref{fig:3dzjet} on the left.

The measurement was performed in the dimuon final state and unfolded simultaneously in all three observables to correct for detector effects.
The triple-differential binning provides enhanced sensitivity to parton distribution functions compared to inclusive or single-differential measurements.

The cross section was measured for $p_T^Z$ values up to $1\,\mathrm{TeV}$ with typical experimental uncertainties of $1.5\%$--$5\%$ in the central phase space, increasing in regions of large $y_b$ or $y^*$ due to limited event yields.
The results were compared to state-of-the-art predictions at next-to-next-to-leading order in QCD, supplemented with electroweak and non-perturbative corrections, and showed overall good agreement, as shown in Fig.~\ref{fig:3dzjet} on the right.

This measurement provides strong constraints on parton distribution functions and demonstrates the power of multi-differential observables for precision QCD studies at the LHC, enhancing sensitivity to the underlying parton dynamics compared to inclusive observables.

\section{Jet mass of boosted W bosons and extraction of $m_W$}

The CMS Collaboration reported a measurement of the jet mass of highly boosted W bosons decaying hadronically~\cite{CMS:SMP-24-012} in proton--proton collisions at $\sqrt{s}=13\,\mathrm{TeV}$, using the full Run~2 dataset corresponding to an integrated luminosity of $138\,\mathrm{fb}^{-1}$.

At large transverse momenta ($p_T > 650\,\mathrm{GeV}$), the decay products of the W boson become highly collimated and are reconstructed as a single large-radius jet.
The characteristic two-prong substructure of these jets is exploited to distinguish them from the dominant background of quark- and gluon-initiated jets.

The jet mass was computed using the soft-drop grooming algorithm, which removes soft, wide-angle radiation and improves the resolution of the mass observable.
The measurement was performed double-differentially as a function of the jet transverse momentum and the groomed jet mass, and was unfolded to the particle level using a maximum-likelihood approach.

The measured distributions were compared with predictions from Monte Carlo event generators including higher-order QCD and electroweak corrections, and were found to be well described within uncertainties.
From these distributions, the W boson mass was extracted in the fully hadronic final state, yielding
$m_W = 80.77 \pm 0.57\,\mathrm{GeV}$.

This result constitutes the first determination of the W boson mass in an all-jets final state at a hadron collider and demonstrates the potential of jet substructure techniques for precision measurements in complex final states at the High-Luminosity LHC.

\section{Summary and Outlook}

A set of recent measurements of W and Z boson production by the ATLAS and CMS Collaborations has been presented, spanning searches for physics beyond the Standard Model and precision studies of perturbative QCD and electroweak dynamics.

A common theme across these results is the increasing role of precision measurements as powerful probes of the Standard Model.
The large datasets collected during Run~2 enable multi-differential measurements and the exploration of regions of phase space that were previously statistically limited, significantly enhancing sensitivity to parton distribution functions and higher-order effects.

The combination of improved experimental techniques and state-of-the-art theoretical predictions allows for increasingly stringent tests of the Standard Model.
Remaining differences between data and predictions provide valuable input for further refinements of theoretical calculations and Monte Carlo modelling.

With the larger datasets from Run~3 and the future High-Luminosity LHC, substantial gains in precision and kinematic reach are expected, enabling even more detailed studies and further strengthening the interplay between experiment and theory.
In addition, dedicated low pile-up data-taking at the end of Run~3, expected to provide about $1\,\mathrm{fb}^{-1}$ of integrated luminosity, will be crucial for measurements that are currently limited by the statistical precision of the available low pile-up dataset.

\section*{References}
\bibliography{moriond}


\end{document}